# Ferroelectric Dead Layer Driven by a Polar Interface

Y. Wang,[1] M. K. Niranjan,[1] K. Janicka,[1] J. P. Velev,[2] M. Ye. Zhuravlev,[3]

S. S. Jaswal,[1] and E. Y. Tsymbal [1]

[1] *Department of Physics and Astronomy, Nebraska Center for Materials and Nanoscience,*

*University of Nebraska, Lincoln, Nebraska 68588, USA*

[2] *Department of Physics and Institute for Functional Nanomaterials,*

*University of Puerto Rico, San Juan, Puerto Rico 00931, USA*

[3] *Kurnakov Institute for General and Inorganic Chemistry, Russian Academy of Sciences,*

*119991 Moscow, Russia*

## Abstract

Based on first-principles and model calculations we investigate the effect of polar interfaces on the ferroelectric stability of thin-film ferroelectrics. As a representative model, we consider a $TiO_2$-terminated $BaTiO_3$ film with LaO monolayers at the two interfaces that serve as doping layers. We find that the polar interfaces create an intrinsic electric field that is screened by the electron charge leaking into the $BaTiO_3$ layer. The amount of the leaking charge is controlled by the boundary conditions which are different for three heterostructures considered, namely Vacuum/LaO/$BaTiO_3$/LaO, LaO/$BaTiO_3$, and $SrRuO_3$/LaO/$BaTiO_3$/LaO. The intrinsic electric field forces ionic displacements in $BaTiO_3$ to produce the electric polarization directed into the interior of the BaTiO$_3$ layer. This creates a ferroelectric dead layer near the interfaces that is non-switchable and thus detrimental to ferroelectricity. Our first-principles and model calculations demonstrate that the effect is stronger for a larger effective ionic charge at the interface and longer screening length due to a stronger intrinsic electric field that penetrates deeper into the ferroelectric. The predicted mechanism for a ferroelectric dead layer at the interface controls the critical thickness for ferroelectricity in systems with polar interfaces.

# 1. Introduction

Ferroelectric materials that are characterized by a switchable macroscopic spontaneous polarization have attracted significant interest due to their technological applications in ferroelectric field-effect transistors and nonvolatile random access memories. [1-3] To increase the capacity of the storage media, it becomes essential to bring ferroelectricity into the nanometer scale. Much experimental and theoretical effort has been devoted to determine the critical thickness of ferroelectric thin films and elucidate its origin. Based on early experiments it was believed that the ferroelectricity vanishes below a critical thickness of a few tens of nm [4] due to the depolarizing field produced by polarization charges on the two surfaces of the ferroelectric film. [5] There is a depolarizing field in a ferroelectric film placed between two metal electrodes due to incomplete screening which is inversely proportional to the thickness of the ferroelectric. As a result, there is a critical thickness for a ferroelectric below which depolarizing field becomes too large resulting in the suppression of ferroelectricity. Recent experimental and theoretical findings demonstrate, however, that ferroelectricity persists down to a nanometer scale. [6] In particular, it was discovered that in organic ferroelectrics ferroelectricity can be sustained in thin films of monolayer thickness. [7] In perovskite ferroelectric oxides ferroelectricity was observed in nm-thick films. [8-13] These experimental results are consistent with first-principles calculations that predict that the critical thickness for ferroelectricity in perovskite films can be as small as a few lattice parameters. [14-19] The existence of ferroelectricity in ultrathin films opens possibilities for novel nanoscale devices, such as ferroelectric [20-26] and multiferroic [27-30] tunnel junctions.

Ferroelectric properties of thin films placed between two metal electrodes to form a ferroelectric capacitor or a ferroelectric tunnel junction are affected by a number of factors. It was demonstrated that in addition to the screening associated with free charges in the metal electrodes [5] there is an important contribution resulting from ionic screening if electrodes are oxide metals, such as $SrRuO_3$. [17] It was also predicted that the interface bonding at the ferroelectric-metal interfaces influences strongly the ferroelectric state through the formation of intrinsic dipole moments at the interfaces, as determined by the chemical constituents and interfacial metal-oxide bonds. [16] For some interfaces, these dipole moments are switchable and may enhance the ferroelectric instability of the thin film, which is interesting for engineering the electrical properties of thin-film devices. [31] For other interfaces, however the effect of interface bonding is detrimental and leads to the "freezing" of polar displacements in the interfacial region, thus resulting in a ferroelectrically inactive layer near the interface. [16]

In this paper, we explore the effect of a *polar* interface on the ferroelectric stability of a thin film of $BaTiO_3$ and the formation of a ferroelectric dead (non-switchable) region near the interface. This study in motivated by our recent work devoted to the control of two-dimensional electronic gas (2DEG) properties at oxide interfaces through ferroelectric polarization. [32,33] A discovery of a 2DEG at oxide interfaces between insulating perovskite $SrTiO_3$ and $LaAlO_3$ has attracted significant interest. [34] The 2DEG has a very high carrier density and a relatively high carrier mobility making it attractive for applications in nanoelectronics. [35,36] At low temperatures the oxide 2DEG may become magnetic [37] or superconducting [38]. These interesting properties of the 2DEG at oxide interfaces have stimulated significant research activity both in experiment [39-47] and in theory [48-60].

Using a ferroelectric material to form a 2DEG is interesting due to a new functionality that allows controlling 2DEG properties. We have predicted that the switching of spontaneous ferroelectric polarization allows modulations of the carrier density and consequently the conductivity of the 2DEG. [32,33] This effect occurs due to the screening charge at the interface that counteracts the depolarizing electric field and depends on polarization orientation. A necessary condition for such a new functionality is the ability of a ferroelectric polarization to switch involving a nm-thick region adjacent to the interface where the 2DEG is formed. [53] However, the polar interface which is required for producing a 2DEG [61] may affect the ferroelectric stability due to the electric field associated with the polar interface. This intrinsic electric field is determined by the electron charge distributed near the interface to screen the ionic charge of the interfacial plane and penetrates into the ferroelectric. The intrinsic electric field is detrimental to ferroelectricity due to pinning ionic displacements near the interface that prevents their switching. These arguments indicate the important role of polar interfaces in ferroelectric stability and serve as the motivation for the present study. The importance of this issue also follows from the fact that polar interfaces may occur in ferroelectric capacitors and tunnel junctions. For example, $A^{1+}B^{5+}O_3$ perovskite ferroelectrics (such as $KNbO_3$) have alternating charged monolayers of $(AO)^-$ and $(BO_2)^+$ along the [001] direction so that the (001) surface is expected to be polar. In addition, metal oxide electrodes may have charge uncompensated atomic planes along the growth direction (such as La$_{2/3}$Sr$_{1/3}$MnO$_3$(001)), so that epitaxially grown films may produce polar interfaces.

## 2. Present investigations

To investigate the effect of polar interfaces on the ferroelectric stability, as representative systems, we consider three different types of heterostructures: (1) Vacuum/LaO/BaTiO$_3$/LaO, (2) LaO/BaTiO$_3$, and (3) SrRuO$_3$/LaO/BaTiO$_3$/LaO. In all the three systems a LaO monolayer at the interface with a TiO$_2$-terminated BaTiO$_3$ produces a *polar interface* and serves as a doping layer donating an electron at the interface that leaves behind a positively charged (LaO)$^+$ monolayer. This charge pins the Fermi energy close to the bottom of the conduction band of BaTiO$_3$ and may penetrate ("leak") into BaTiO$_3$ producing a band bending on a scale of the screening length. Similar to the results in Ref 53, we find that the conduction band minimum (CBM) in the middle monolayer of a 21.5 unit cell BaTiO$_3$ is approximately 0.2eV above the Fermi energy and substantial band bending is seen resulting in the CBM below the Fermi energy only in the layers close to the interface (see Appendix I). For the Vacuum/LaO/BaTiO$_3$/LaO structure, it is expected that the charge equal to –$e$ per a lateral unit cell leaks from the interface into the BaTiO$_3$ layer. For the LaO/BaTiO$_3$ system, as follows from symmetry, a charge equal to –0.5$e$ per lateral unit cell leaks into BaTiO$_3$. For the SrRuO$_3$/LaO/BaTiO$_3$/LaO structure, a positive charge of the (LaO)$^+$ monolayer is screened within the SrRuO$_3$ metal electrode leading to about –0.26$e$ charge leaking into BaTiO$_3$ (as follows from our first-principles calculation discussed below). Therefore, the three systems considered are different by the amount of electron charge penetrating into BaTiO$_3$.

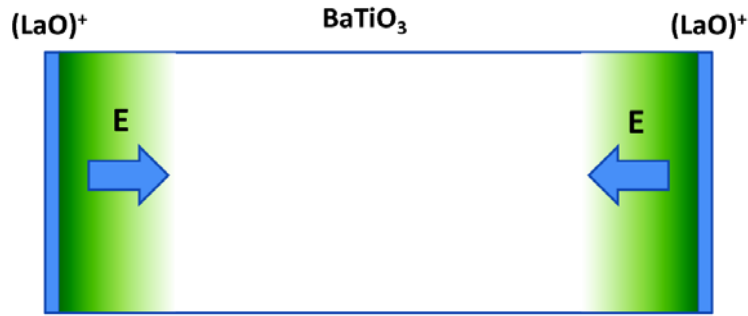

**Fig. 1:** (color online) Charge distribution and the intrinsic electric field in a BaTiO$_3$ film terminated with positively charged (LaO)$^+$ monolayers at the left and right interfaces. A gradient shadow region indicates the screening charge penetrating into BaTiO$_3$. Blue arrows indicate the intrinsic electric field that has a tendency to pin the local polarization along the field direction.

When describing the electron distribution within the BaTiO$_3$ it is convenient to introduce the notion of *effective ionic charge* which is a sum of the positive charge at the LaO monolayer and a negative electron charge distributed on the other side of the interface. Thus, the screening charge in BaTiO$_3$ is associated with the effective ionic charge, which according to the charge conservation condition is equal to +$e$, +0.5$e$, and +0.26$e$ per lateral unit cell area for the three systems respectively. The interface ionic charge creates an *intrinsic electric field* pointing into the BaTiO$_3$ layer as shown in Fig. 1. This field is screened by the negative charge within the screening length of about 1 nm from the interface. Within this region the intrinsic electric field forces ionic displacements in BaTiO$_3$ to produce the electric dipole moments directed into the interior of the BaTiO$_3$ layer. The intrinsic electric field is not related to the polarization charge but pins the electric polarization in the direction away from the interface. This creates a *ferroelectric dead layer* near the interfaces that is non-switchable and thus detrimental to ferroelectricity. The effect is stronger for a system, in which the effective ionic charge at the interface is larger. A longer screening length produces a lesser screening effect making the electric field to penetrate deeper into the ferroelectric layer and thus producing a stronger pinning effect. These predictions are elaborated in detail using first-principles and model calculations as discussed below.

## 3. Systems and theoretical methods

To investigate the effect of polar interfaces on the ferroelectric stability we employ first-principles calculations based on density functional theory [62,63] along with a model approach based on the Landau-Ginzburg-Devonshire theory. [64,65] As was discussed above we consider three types of superlattices: Vacuum/LaO/(BaTiO$_3$)$_m$/LaO, LaO/(BaTiO$_3$)$_n$, and (SrRuO$_3$)$_{5.5}$/LaO/(BaTiO$_3$)$_m$/LaO, where *m* and *n* denote the number of unit cells of BaTiO$_3$: $m = 8.5$ and $n = 8.5$, 14.5 or 21.5. All the layered heterostructures are stacking in the (001) planes with the z-axis pointing normal to the planes in the [001] direction. We impose periodic boundary conditions and assume that BaTiO$_3$ is TiO$_2$-terminated at both interfaces. In our structural model we include implicitly a SrTiO$_3$ substrate by constraining the in-plane lattice constants of the supercell structures to the calculated bulk lattice constant of SrTiO$_3$, *i.e.* a = 3.871 Å. This constraint produces a strain on bulk BaTiO$_3$ resulting in the tetragonal distortion of c/a = 1.04 and ferroelectric polarization of 0.26 C/m$^2$, as calculated using the Berry's phase method. [66]

Density-functional calculations are performed using the projected augmented wave (PAW) method [67] implemented within the Vienna Ab-Initio Simulation Package (VASP). [68] The exchange-correlation effects are treated within the local density approximation (LDA). The electron wave functions are expanded in a plane-wave basis set limited by a cutoff energy of 500eV. Calculations are performed using the 8×8×1 Monkhorst-Pack k-point mesh. [69] The self-consistent calculations are converged to 10$^{-5}$ eV/cell. Ferroelectric states are obtained by relaxing all the ions in the superlattices starting with the displacement pattern of the bulk tetragonal soft mode with polarization pointing perpendicular to the planes until the forces on the ions are less 20meV/Å.

The goal of using the phenomenological model is to investigate the influence of the intrinsic electric field $E_i$ at the polar interface (Fig. 1) on ferroelectric polarization. Within the Landau-Ginzburg-Devonshire theory of a ferroelectric film, [64, 65] the free energy is expressed as follows: [70, 71]

$$F = \int (\frac{A}{2}P^2 + \frac{B}{4}P^4 + \frac{C}{2}(\nabla \cdot P)^2 - E_i P - \frac{1}{2}E_{ds}P)dV + \int \frac{C}{2\delta}\{[P(0)+P_i]^2 + [P(d)-P_i]^2\}dS \quad . \quad (1)$$

Here $P(z)$ is a ferroelectric polarization, which is assumed to be homogeneous in the *x-y* plane of the film, $P_i$ is the interface polarization, and $E_i$ is the intrinsic electric field not related to the polarization. The last term in the volume integral represents the self-energy of the depolarizing field $E_{ds}$ that includes screening (see Appendix II). Constant $A$ is given by Curie-Weiss law $A = \dfrac{T - T_C}{C_0 \varepsilon_0}$, where $C_0$ and $T_C$ are the Curie constant and temperature, and $\varepsilon_0$ is the permittivity of free space. Constant $B$ is given by $B = -A/P_0^2$, where $P_0$ is the polarization at zero temperature. Constant $C$ is related to $A$ as follows: $C = -Aa_0^2$, where $a_0$ is of the order of lattice spacing. [70] The extrapolation length $\delta$ is a phenomenological parameter associated with the derivatives of the $P(z)$ at the interface.

We obtain the intrinsic electric field $E_i$ and depolarizing field $E_{ds}$, as described in the Appendix, assuming that the free charge that screens the effective ionic charge at the interfaces decays exponentially into the ferroelectric layer with decay length $\lambda$ and is redistributed between the interfaces to screen the depolarizing field. Variation of the free energy functional given by Eq. (1) over the polarization yields the Euler–Lagrange equation for polarization

$$C \frac{d^2 P}{dx^2} = AP + BP^3 - E_i + \frac{1}{2} \frac{2P - R\overline{P} - \overline{RP}}{\varepsilon_f} \qquad (2)$$

subject to boundary conditions

$$\left( P \mp \delta \frac{dP}{dz} \right)_{z=0,d} = \mp P_i . \qquad (3)$$

In Eq. (2) $\bar{P}$ and $\overline{RP}$ are the average polarization $P(z)$ and the average function $R(z)P(z)$ over film thickness, where $R(z)$ is given by eq. (A.8). Eq. (2) can be solved numerically using an iterative procedure. Numerical calculations suggest that the last term in Eq. (2) can be replaced with a reasonable accuracy by $\dfrac{P-R\bar{P}}{\varepsilon_f}$ which simplifies the convergence to a self-consistent solution. In particular calculations, we use a set of parameters appropriate for BaTiO$_3$: $T_C = 403 K$, $P_0 = 0.27 C/m^2$, $C_0 = 1.7 \cdot 10^5$, [72] and $a_0 = 0.8 nm$. [73,74] The dielectric permittivity $\varepsilon_f$ originates from the non-ferroelectric lattice modes [75] and for BaTiO$_3$ is assumed to be $\varepsilon_f = 90\varepsilon_0$. [25] This value describes adequately the depolarizing field in BaTiO$_3$ obtained from first-principles calculations for a SrRuO$_3$/BaTiO$_3$/SrRuO$_3$ Ferroelectric Tunnel Junction. [29] In the calculations we use different effective ionic charges $\sigma_i$ at the BaTiO$_3$ surfaces which enter eq. (A.2) for the intrinsic electric field. For the Vacuum/LaO/BaTiO$_3$/LaO system, the effective interface charge $\sigma_i$ is assumed to be $+e$ per lateral unit cell area, while for LaO/BaTiO$_3$ and SrRuO$_3$/LaO/BaTiO$_3$/LaO systems lattices they are assumed to be $+0.5e$ and $+0.26e$ respectively. Other parameters, i.e. $\lambda$, $\delta$ and $P_i$, are the adjustable parameters of the model.

### 4. Electronic structure

First, we analyze the electronic band structure of Vacuum/LaO/(BaTiO$_3$)$_{8.5}$/LaO, LaO/(BaTiO$_3$)$_{8.5}$, and (SrRuO$_3$)$_{5.5}$/LaO/(BaTiO$_3$)$_{8.5}$/LaO heterostructures. In all the structures the presence of the interfacial LaO donor monolayer produces an extra valence charge that resides near the interface, partly or fully leaking into the BaTiO$_3$ layer. The latter fact is evident from Fig. 2, which shows the density of states (DOS) projected onto the 3d-orbitals of Ti atoms located at the left and right interfaces (solid lines) and in the middle of the $BaTiO_3$ layer (shaded area) for the three heterostructures. There are occupied states at and below the Fermi level (placed at zero energy in Fig. 2), the density of which decreases with the distance from the interface in the interior of $BaTiO_3$. The attenuation length depends on the state energy in the band gap. Typically the shortest decay length corresponds to energies close to the middle of the gap. At the band gap edges, however, adjacent to the conduction band minimum or the valence band maximum the decay length diverges due to the imaginary part of the complex wave vector tending to zero. [76] The average decay length is about 1nm for all the three heterostructures, which is consistent with the calculation for the $LaAlO_3/SrTiO_3$ system. [53] We note the DOS at the Fermi energy decays at a larger scale and involves metal-induced gap states, [53] so that a much larger thickness of $BaTiO_3$ is required to approach a zero DOS. This is discussed in detial in Appendix I where we analyze the band alignment for the $LaO/(BaTiO_3)_{21.5}$ system. The total leakage charge can be calculated by integrating the total DOS lying within the band gap of $BaTiO_3$ upto the Fermi energy. We find for Vacuum/LaO/$(BaTiO_3)_{8.5}$/LaO, LaO/$(BaTiO_3)_{8.5}$, and $(SrRuO_3)_{5.5}$/LaO/$(BaTiO_3)_{8.5}$/LaO heterostructures that this charge is equal to 1$e$, 0.5$e$, and 0.26$e$ per unit cell area respectively.

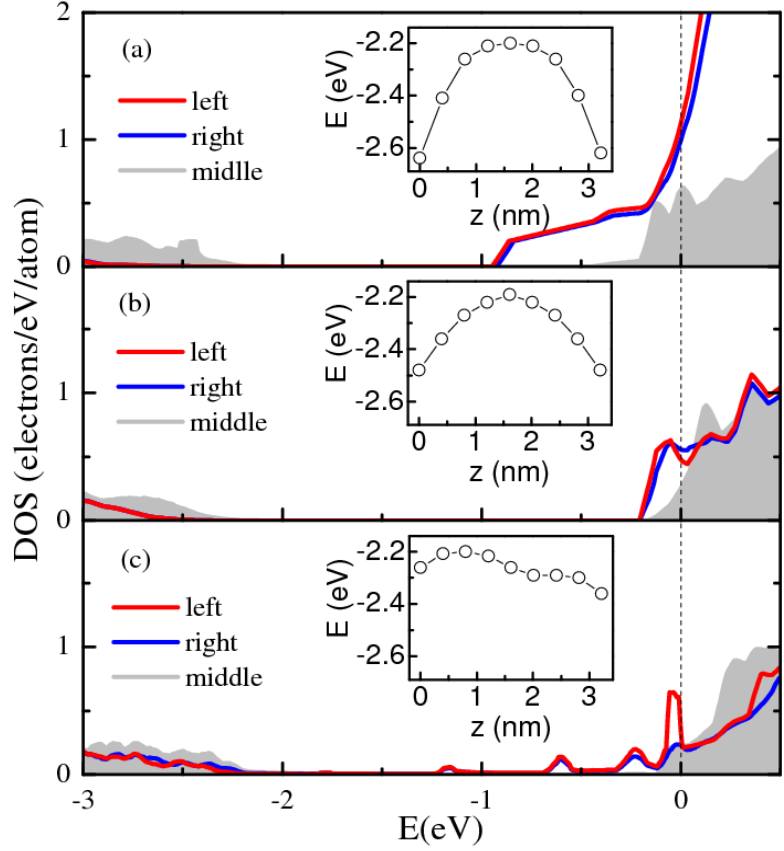

**Fig. 2:** Density of states (DOS) projected onto the $3d$ states of Ti atoms located at the left and right interfaces (solid lines) and in the middle of the BaTiO$_3$ layer (shaded area) for (a) Vacuum/LaO/(BaTiO$_3$)$_{8.5}$/LaO, (b) LaO/(BaTiO$_3$)$_{8.5}$, and (c) (SrRuO$_3$)$_{5.5}$/LaO/(BaTiO$_3$)$_{8.5}$/LaO heterostructures. The Fermi level lies at zero energy and denoted by the dashed line. The insets show the valence band maximum (VBM) with respect to the Fermi energy within the BaTiO$_3$ layer as a function of $z$.

As is evident from the DOS plotted in Fig. 2a, for a Vacuum/LaO/(BaTiO$_3$)$_{8.5}$/LaO heterostructure, there is a significant shift of the valence band maximum (VBM) in the middle of BaTiO$_3$ layer as compared to the interfaces. This is an indication of the electrostatic potential change in the insulator and hence an intrinsic electric field. We show the variation of the electrostatic potential energy across the BaTiO$_3$ layer in the inset of Fig. 2a by plotting the VBM energy (normalized to zero at maximum) as a function of position $z$ (measured in unit cells) within BaTiO$_3$. A significant increase in the electrostatic potential energy at the two interfaces that is seen from the inset indicates the presence of the electric field pointing away from the interfaces. The magnitude of this field close to the interfaces is about 0.5 V/nm. Almost symmetric variation of the electrostatic potential is the indication of the absence of ferroelectric polarization in BaTiO$_3$ which is expected to break the symmetry.

A similar behavior is present in a LaO/(BaTiO$_3$)$_{8.5}$ heterostructure. In this case, however, as seen from the inset in Fig. 2b, the change in the electrostatic energy and hence the electric field are reduced by a factor of two, as compared to the Vacuum/LaO/(BaTiO$_3$)$_{8.5}$/LaO system. This reduction is the consequence of the reduced value of the electric charge per interface (0.5$e$ versus 1$e$) that screens the (LaO)$^+$ and penetrates into BaTiO$_3$. The electric field is still sufficiently large to not allow for ferroelectric polarization to develop for the 8.5 unit cell thick BaTiO$_3$ layer, as follows from the almost symmetric energy profile seen in the inset of Fig. 2b.

The situation changes for a (SrRuO$_3$)$_{5.5}$/LaO/(BaTiO$_3$)$_{8.5}$/LaO heterostructure, where the large portion of the screening charge resides in the SrRuO$_3$ electrodes allowing only 0.26 $e$/interface to leak into BaTiO$_3$. The reduced value of the charge and hence the intrinsic electric field makes it possible for ferroelectric polarization to develop in the system, as we will see below. An indirect indication of this fact is a complex electrostatic energy profile across the BaTiO$_3$ layer (see the inset in Fig. 2c) which in addition to the intrinsic electric field includes contributions from the depolarizing field created by a non-uniform polarization and the associated screening charge.

We note that the amount and the penetration depth of the electron charge into BaTiO$_3$ are largely determined by the position of the Fermi energy with respect to the bottom of the conduction band of BaTiO$_3$. The latter is controlled by the conduction band bending due to the electrostatic potential associated with the screening charge that places the conduction band minimum at a certain position above the Fermi energy. Therefore, the well-known problem of LDA to underestimate the band gap of insulator and predict an incorrect band offset between metal and insulators does not affect significantly the predicted results.

## 5. Atomic structure and polarization

The effect of the intrinsic electric field at the interface on ferroelectric polarization is evident from polar atomic displacements in the studied systems which are correlated with a ferroelectric instability. Figs. 3a,b,c show the displacements of cations (Ba and Ti) with respect to oxygen anions in the BaTiO$_3$ layer for Vacuum/LaO/(BaTiO$_3$)$_{8.5}$/LaO, LaO/(BaTiO$_3$)$_{8.5}$, and (SrRuO$_3$)$_{5.5}$/LaO/(BaTiO$_3$)$_{8.5}$/LaO heterostructures respectively. As seen from Fig. 3a, the displacements profile is nearly inversion-symmetric and, although the displacements are very large (about 0.2 Å) close to the interfaces, they have opposite sign. This behavior is the consequence of the intrinsic electric field pointing away from the interfaces (Fig. 1) that forces the polarization to be pinned in the same direction. The LaO/(BaTiO$_3$)$_{8.5}$ system exhibits a similar displacements profile, although the magnitude of the displacements is somewhat reduced at the interface (Fig. 3b). The latter fact is an indication of the reduced intrinsic electric field at the interface due to a reduced charge penetrating into BaTiO$_3$ near the interfaces, as was discussed above. For the (SrRuO$_3$)$_{5.5}$/LaO/(BaTiO$_3$)$_{8.5}$/LaO system the intrinsic electric field is further reduced that allows a ferroelectric polarization to develop. This follows from the asymmetric polarization profile in Fig. 3c with a larger portion of electric dipoles pointing in the positive direction along the *z* axis (i.e. from left to right interface).

We estimate the local polarization distribution within BaTiO$_3$ using a model based on the Born effective charges. [77] For this purpose we compute the local polarization *P*(*z*) by averaging over the *z*-dependent displacements in the BaTiO$_3$ primitive unit cell using equation $P(z) = \frac{e}{\Omega} \sum_{m=1}^{N} Z_m^* \delta u_m$. Here *N* is the number of atoms in the primitive unit cell, $\delta u_m$ is the change of the position vector of the *m*-th atom, and $\Omega$ is the volume of the unit cell. The Born effective charges $Z_m^*$ are 2.77, 7.25, -2.15, -5.71 for Ba, Ti, O$_\perp$ and O$_\parallel$ ions respectively. [78] Using these values the polarization of the strained bulk BaTiO$_3$ is calculated to be 0.27/m$^2$, which is in excellent agreement with our calculation based on the Berry phase method, i.e. 0.26 C/m$^2$. We note that the method based on the Born effective charges calculated for bulk ferroelectrics cannot provide a quantitatively accurate description of the local polarization distribution in heterostructures due to the effects of interfaces and local fields which do not exist in the bulk.

Nevertheless, we find this approach valuable for a semi-quantitative estimate of the polarization behavior and comparison with our phenomenological model (see sec. 6).

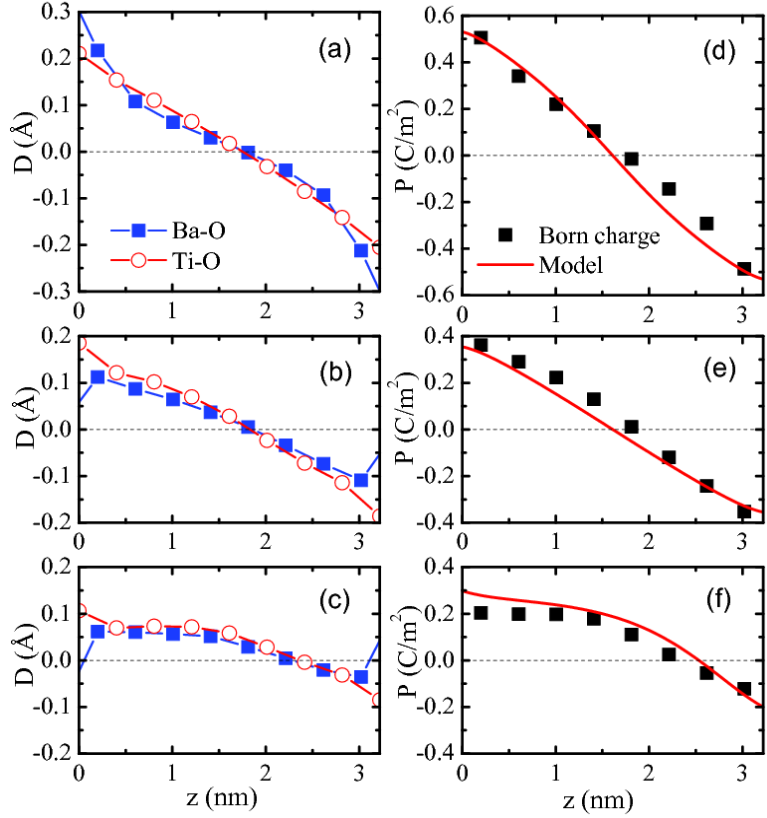

**Fig. 3:** Displacements (D) of cations (Ba and Ti) with respect to oxygen anions (a, b, c) and local polarizations (P) (d, e, f) across a BaTiO$_3$ layer in Vacuum/LaO/(BaTiO$_3$)$_8$/LaO (a, d), LaO/(BaTiO$_3$)$_{8.5}$ (b, e) and (SrRuO$_3$)$_{5.5}$/LaO/(BaTiO$_3$)$_{8.5}$/LaO (c, f) heterostructures. Solid squares and open circles in (a-c) denote Ba-O and Ti-O displacements respectively. In figures (d-f) the local polarizations are obtained using the displacements calculated from first-principles in conjunctions with the Born effective charges (squares) and from the phenomenological model (solid lines).

The results are displayed in Figs. 3d-f (squares). It is seen that for Vacuum/LaO/(BaTiO$_3$)$_{8.5}$/LaO (Fig. 3d) and LaO/(BaTiO$_3$)$_{8.5}$ (Fig. 3e) heterostructures the polarization profiles are nearly inversion-symmetric resulting in very low average polarization values: 0.029 and 0.037 C/m$^2$ for the two systems respectively. For the (SrRuO$_3$)$_{5.5}$/LaO/(BaTiO$_3$)$_{8.5}$/LaO heterostructure (Fig. 3f) the ferroelectric polarization is more pronounced (the average value is 0.09 C/m$^2$) due to the screening of the depolarizing field in the SrRuO$_3$ electrodes. Nevertheless, even in this case the presence of the intrinsic electric field associated with the polar interfaces force the dipole moments in BaTiO$_3$ to be pointed away from the interfaces, resulting in a ferroelectric dead layer as discussed below.

The presence of a ferroelectric dead layer, i.e. the BaTiO$_3$ region near the interface that does not switch upon ferroelectric polarization reversal, can be seen from the dependence of polar displacements and local polarization distribution as a function of BaTiO$_3$ thickness. To study this dependence we have performed calculations for LaO/(BaTiO$_3$)$_n$ heterostructures containing $n$ = 8.5, 14.5 or 21.5 unit cells of BaTiO$_3$ that correspond to BaTiO$_3$ thicknesses $t$ = 3.2, 5.6, 8.4 nm. Since the boundary conditions at the two interfaces are identical independent of BaTiO$_3$ thickness this calculation reveals the effect of polar interfaces on the ferroelectric polarization. The results are displayed in Figs. 4a-f. It is seen that as the BaTiO$_3$ thickness increases, the ferroelectric polarization becomes more stable involving a larger thickness of the BaTiO$_3$ layer. The average polarization of BaTiO$_3$ increases from $\bar{P}$ = 0.029 C/m$^2$ for $t$ = 3.2 nm to $\bar{P}$ = 0.095 C/m$^2$ and 0.12 C/m$^2$ for $t$ = 5.6 and 8.4 nm respectively. It is notable, however, that even at a relatively large thickness of 8.4 nm the ferroelectric polarization in the middle of the BaTiO$_3$ layer is 0.21 C/m$^2$, i.e. it does not reach the respective bulk value 0.27 C/m$^2$.

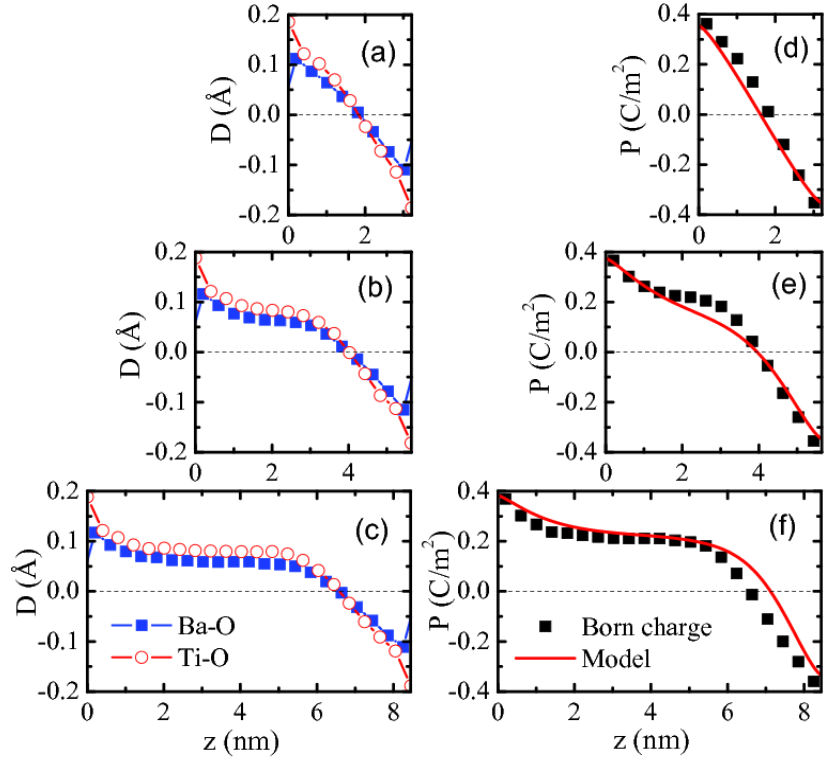

**Fig. 4:** Displacements (D) of cations (Ba and Ti) with respect to oxygen anions (a, b, c) and local polarizations (P) (d, e, f) across the BaTiO$_3$ layer in LaO/(BaTiO$_3$)$_n$ heterostructures containing $n$ = 8.5 (a, d), 14.5 (b, e) or 21.5 (c, f) unit cells of BaTiO$_3$. Solid squares and open circles in (a-c) denote Ba-O and Ti-O displacements respectively. In figures (d-f) the local polarizations are obtained using the displacements calculated from first-principles in conjunctions with the Born effective charges (squares) and from the phenomenological model (solid lines).

The enhancement of polarization at large BaTiO$_3$ thickness does not affect significantly the BaTiO$_3$ region adjacent to the right interface where the polar displacements remain opposite to the spontaneous polarization displacements. From comparison of Figs. 4d-f it is seen that the region where the polar displacements are reversing is almost independent of BaTiO$_3$ thickness and cover about 3 nm thicknesses near the interface. This may be regarded as a ferroelectric dead layer and determines the critical thickness for ferroelectricity in this system.

We note that in the previous calculation of critical thickness for ferroelectricity of a KNbO$_3$ film placed between two metal electrodes [16] the effect of the polar interfaces may also play a role due to the KNbO$_3$(001) layer being comprised of the charged (KO)$^-$ and (NbO$_2$)$^+$ planes along the [001] direction. The predicted ferroelectric domain wall occurring in the SrRuO$_3$/KNbO$_3$ heterostructure may partly be caused by an intrinsic electric field occurring at the polar interface in this system.

## 6. Phenomenological model

The model based on the Landau-Ginzburg-Devonshire theory, as described in sec. 3, was used to further elucidate the effect of polar interfaces on ferroelectric polarization. By explicitly including the effective interface charge and the associated screening charge in the simulation we have fitted the local polarizations obtained from the first-principle calculations and the model based on the Born effective charges. To reduce the number of fitting parameters we fixed $\lambda = 1$nm consistent with our DFT results and $\delta = 1.2$nm. The value of $P_i = -0.45$C/m$^2$ was fitted and was assumed to be the same for all the structures considered. Different effective ionic charges $\sigma_i$ at the BaTiO$_3$ surfaces were used, as described in sec. 3. Figs. 3e-f and 4e-f show results of the fitting by solid lines, demonstrating that the phenomenological model is capable of describing all the major features in the distributions of the local polarization that are obtained from the first-principles displacements combined with the Born effective charges. In particular, the model clearly indicates that the intrinsic electric field associated with polar interfaces pins the atomic displacements near the interfaces and thus is detrimental to ferroelectricity. For a given decay length $\lambda$ inside the ferroelectric, the larger effective ionic charge at the interface creates a wider ferroelectric dead layer near the interface and thus produces a stronger destructive effect on ferroelectricity.

To obtain a further insight into the effect of the intrinsic electric field on ferroelectricity we have modeled the average polarization of a ferroelectric layer terminated by a polar interface allowing the decay length $\lambda$ to be a variable parameter. In the simulation we kept all the other parameters to be the same as those for the LaO/(BaTiO$_3$)$_n$ heterostructure. Fig. 5a shows the resulting average polarization $\bar{P}$ as a function of $\lambda$ for films of different thicknesses $t$. It is seen that with increasing decay length the average polarization of the film decreases and at some value of $\lambda$ it vanishes even for a very thick BaTiO$_3$ layer. This detrimental effect on ferroelectricity originates from the increasing penetration depth of the intrinsic electric field into the ferroelectric layer that broadens a ferroelectrically dead region near the interfaces. On the other hand, when $\lambda$ tends to zero the pinning electric field near the interfaces vanishes and the average polarization approaches its bulk value as the thickness of BaTiO$_3$ is increasing.

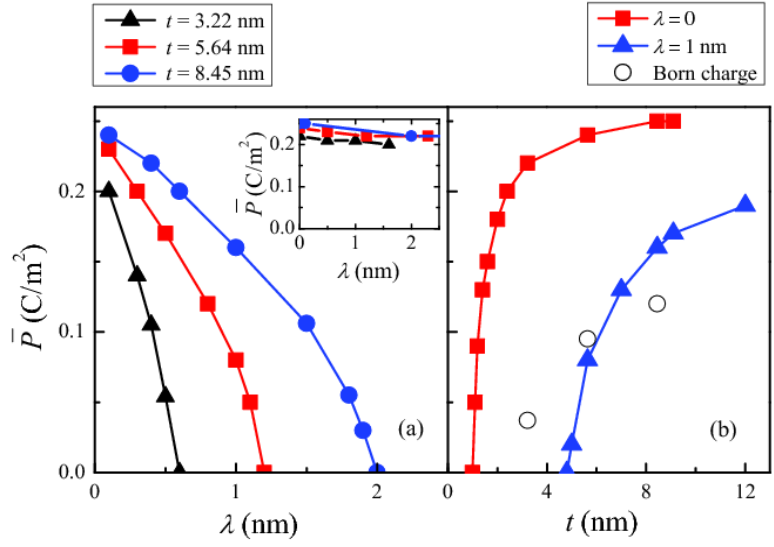

**Fig. 5:** Average polarization $\bar{P}$ of a BaTiO$_3$ film (a) as a function of the decay length $\lambda$ for three film thicknesses $t$ and (b) as a function of the film thickness $t$ for two values of $\lambda$ calculated within the phenomenological model (solid symbols) and using first-principles calculations and the Born charge model (open symbols). The inset in (a) shows $\bar{P}$ for $E_i = 0$.

The stability of ferroelectric polarization strongly depends on the screening length $\lambda$. The latter determines the penetration depth of the screening charge and consequently the electric field into the ferroelectric layer. Increasing the screening length leads to the increase of the critical thickness for ferroelectricity. This statement is confirmed in our model calculations displayed in Fig. 5b that show the average polarization as a function of ferroelectric layer thickness $t$ for different $\lambda$. In this calculation the magnitude of the charge that screens polarization is kept fixed so that only the penetration depth of the charge that screens the intrinsic electric field is varied. It is seen that the increase of the screening length from $\lambda = 0$ to $\lambda$

= 1nm leads to the increase in the critical thickness from about 1 nm to about 5 nm. The latter is qualitatively consistent with our first-principles calculations combined with the model based on the effective Born charges.

We would like to emphasize that the predicted suppression of polarization in our systems is entirely caused by the polar interfaces and the associated intrinsic electric field resulting in a ferroelectric dead layer. Other possible mechanisms detrimental to ferroelectricity such as insufficient screening of polarization charge and the interface bonding do not play a decisive role. The polarization screening due to the redistribution of the free charge between interfaces (see Appendix) is sufficient to maintain the polarization. This fact follows from the calculation we performed within the phenomenological model in which the intrinsic electric field was artificially set equal to zero, i.e. $E_i = 0$, in the free energy given by Eq. (1), but other parameters of the model were kept fixed. The calculation of the average polarization with respect to film thickness $t$ predicts no decay of polarization with $\lambda$ (see the inset in Fig. 5a) and the critical thickness for ferroelectricity of about 1 nm (see Fig. 5b) consistent with previous first-principles results [14-17]. We can also rule out the effect of interface bonding as a possible mechanism of suppression of ferroelectricity. If it were due to the interface bonding, then the ferroelectric polarization in the three structures considered would be similar due to the same LaO monolayers terminating BaTiO$_3$. However, the results of our first-principles calculations suggest a very different behavior for the three systems.

Our results are consistent with the stability of ferroelectric KNbO$_3$ in the presence of 2DEG at the KNbO$_3$/SrTiO$_3$ interfaces predicted earlier. [32, 33] In this case the positive charge at the polar interface is screened by a free electron charge penetrating almost equally both to the KNbO$_3$ and SrTiO$_3$ layers. As follows from symmetry, this implies that only 0.25$e$ per interface leak into the KNbO$_3$ with decay length of about 1nm. Due to the same leakage charge and the decay length, the local polarization distribution is expected to be qualitatively similar to that found for the SrRuO$_3$/LaO/BaTiO$_3$/LaO heterostructure. This fact is evident from the similarity between displacements shown in Fig. 3c and those in Fig. 4a of ref. [33].

## 7. Summary

Based on first-principles model and calculations we have investigated the effect of polar interfaces on the ferroelectric stability of thin-film ferroelectrics using Vacuum/LaO/BaTiO$_3$/LaO, LaO/BaTiO$_3$, and SrRuO$_3$/LaO/BaTiO$_3$/LaO heterostructures as representative systems. In all the three systems a LaO monolayer at the interface with a TiO$_2$-terminated BaTiO$_3$ produces a polar interface and serves as a doping layer donating an electron at the interface that compensates the ionic charge of the positively charged (LaO)$^+$ monolayer. This interface ionic charge creates an intrinsic electric field at the interface which is screened by the screening electron charge leaking into the BaTiO$_3$ layer within the screening length of about 1 nm from the interface. The three systems considered are different by the amount of the screening charge changing from –$e$ per a lateral unit cell to –0.5$e$ and –0.26$e$ and thus *effective ionic charge* for the Vacuum/LaO/BaTiO$_3$/LaO, LaO/BaTiO$_3$, and SrRuO$_3$/LaO/BaTiO$_3$/LaO systems respectively. Within the screening region the intrinsic electric field forces ionic displacements in BaTiO$_3$ to produce the dipole moments pointing into the interior of the BaTiO$_3$ layer and thus pins the polarization near the interface. This creates a ferroelectric dead layer near the interfaces that is non-switchable and thus detrimental to ferroelectricity. Our first-principles and model calculations demonstrate that the effect is stronger for a system, in which the effective ionic charge at the interface is larger and the screening length is longer resulting in a stronger intrinsic electric field that penetrates deeper into the ferroelectric. The predicted mechanism for a ferroelectric dead layer at the interface controls the critical thickness for ferroelectricity in systems with polar interfaces.

## Appendix I: Band alignment in LaO/(BaTiO$_3$)$_{21.5}$ system

Here we analyze the band alignment in LaO/(BaTiO$_3$)$_{21.5}$ system. Fig. 6a shows the calculated density of states (DOS) on the TiO$_2$ monolayers located at different distances from the LaO monolayers at the interface. A substantial band bending is seen resulting in the change of the valence band maximum (VBM) and the conduction band minimum (CBM) as a function of $l$. This is due to the variation of the macroscopic electrostatic potential which rigidly shifts the bands with respect to the Fermi energy. The macroscopic electrostatic potential is obtained by performing the macroscopic averaging of the microscopic potential obtained from the supercell calculation that allows filtering out microscopic periodic oscillations in the original data. [79,80] Fig. 6b shows the result of this averaging for the LaO/(BaTiO$_3$)$_{21.5}$ superlattice, where the planer-averaged electrostatic potential (thin line) and the potential additionally averaged over the BaTiO$_3$ c-lattice constant (i.e. macroscopically averaged) (thick line) are shown as a function of the position z within the BaTiO$_3$ layer. We find that the variation of the macroscopic electrostatic potential is consistent with the change in the VBM seen in Fig.6a, as is evident from the respective solid curve plotted in accordance to Fig. 6b. This allows us to find the variation of the conduction band minimum (CBM) in BaTiO$_3$ as follows. We add the calculated band gap of the bulk BaTiO$_3$ with the same lattice constant of 2.2eV to the VBM obtained for the TiO$_2$ monolayer in the middle of BaTiO$_3$ layer. This determines the CBM at this site, which appears to be approximately 0.2 eV above the Fermi energy. Then, we obtain the site variation of the CBM by adding the electrostatic potential difference as shown by the solid curve in the conduction band in Fig.6a. It is clearly seen that CBM below the Fermi energy are confined close to the interface layers.

The approach used here is similar to that applied in ref. 53 to prove the importance of evanescent states in controlling the confinement width of 2DEG. As seen from Fig. 6c the variation in the electrostatic potential shown n Fig. 6b is consistent with the position of the core-like O-2s states. This justifies the method used in ref. 53 to analyze the variation of the electrostatic potential based on the O-2s states.

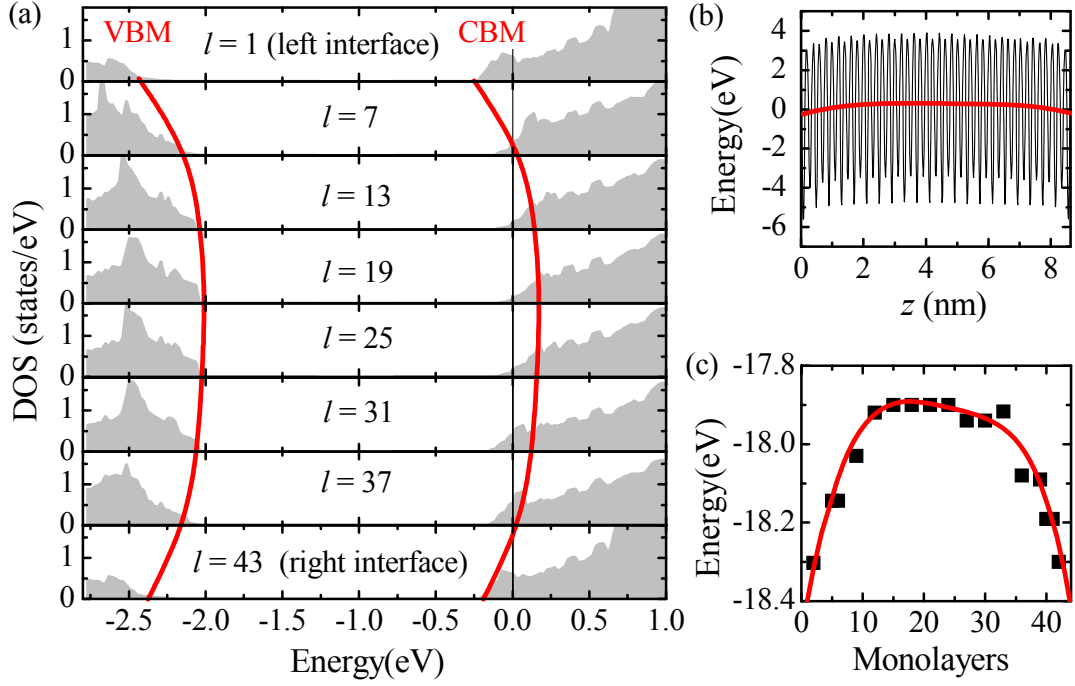

**Fig. 6:** Band alignment in LaO/(BaTiO$_3$)$_{21.5}$ system. (a) Local DOS on TiO$_2$ monolayers ($l$) as a function of the energy. The curved line to the left (right) indicates the position of the valence band maxima (conduction band minima) and the vertical solid line denotes the Fermi energy $E_F$; (b) Planar-averaged electrostatic potential across BaTiO$_3$ (thin line) and the macroscopic electrostatic potential (thick line); (c) Position of the core-like O-2s states for different TiO$_2$ layers (squares) and the macroscopic electrostatic potential shifted by -18.2 eV (solid line).

## Appendix II: Intrinsic and depolarizing fields

Here we calculate the intrinsic electric field $E_i$ associated with the surface ionic charge screened by the free electron charge penetrating into the interior of a BaTiO$_3$ layer (see Fig. 1) and the depolarizing electric field $E_{ds}$ associated with ferroelectric polarization. We consider a ferroelectric thin film of thickness $d$ and polarization $P(z)$ and assume that there is an effective ionic (positive) charge density $\sigma_i$ deposited on each of the two interfaces. As was discuss in section 2, this effective ionic charge is different depending on a particular interface. We assume that there is no polarization and field outside the film. Consistent with the Thomas-Fermi model, we assume that a free electron (negative) charge that screens the ionic charge decays exponentially into the BaTiO$_3$ layer with decay length $\lambda$. Due to charge conservation the resulting volume charge density is given by

$$\rho_i = -\frac{\sigma_i}{\lambda} \frac{e^{\frac{-z}{\lambda}} + e^{\frac{z-d}{\lambda}}}{1 - e^{-\frac{d}{\lambda}}} . \tag{A.1}$$

The resulting electric field produced by both the interface ionic charge and the screening charge is given by:

$$E_i(z) = \frac{\sigma_i}{\varepsilon_f} \frac{e^{\frac{-z}{\lambda}} - e^{\frac{z-d}{\lambda}}}{1 - e^{-\frac{d}{\lambda}}} , \tag{A.2}$$

where $\varepsilon_f$ is the dielectric constant of the ferroelectric at saturation.

The ferroelectric polarization of the film produces the volume polarization charge density $\rho_p = -\frac{dP(z)}{dz}$ and surface polarization charge density at the two interfaces, $\sigma_p(0) = -P(0)$ and $\sigma_p(d) = -P(d)$. The associated (unscreened) depolarizing field is given by

$$E_d(z) = -\frac{P(z)}{\varepsilon_f}, \tag{A.3}$$

where the dielectric permittivity $\varepsilon_f$ originates from the non-ferroelectric lattice modes. [75] This field is partly screened by the redistribution of the free charge density between the two interfaces. Note that this screening of the depolarizing field is different from the screening of the ionic charge located at the interfaces. For simplicity we assume that the screening charge density do not change the shape of the free charge density distribution, but rather adds charge at one interface and removes it at the other interface. Thus the volume screening charge density can be written in the following form:

$$\rho_s = -\frac{\sigma_s}{\lambda} \frac{e^{\frac{-z}{\lambda}} - e^{\frac{z-d}{\lambda}}}{1 - e^{\frac{-d}{\lambda}}}, \tag{A.4}$$

where $\sigma_s$ is the surface screening charge density and the total screening charge (i.e. the integral of (A.4) over film thickness) is equal to zero. The associated screening field is

$$E_s = -\frac{\sigma_s}{\varepsilon_f} \frac{1 + e^{\frac{-d}{\lambda}} - e^{\frac{-z}{\lambda}} - e^{\frac{z-d}{\lambda}}}{1 - e^{\frac{-d}{\lambda}}}. \tag{A.5}$$

In order to obtain the value of $\sigma_s$, we use the short-circuit boundary condition which follow from the periodic boundary condition of the supercell geometry. This condition implies that the electrostatic potential must be equal at the two interfaces, i.e. $\Phi(0) = \Phi(d)$. Using this condition we find

$$\sigma_s = -\overline{P} \frac{1 - e^{\frac{-d}{\lambda}}}{1 + e^{\frac{-d}{\lambda}} - \frac{2\lambda}{d} + \frac{2\lambda}{d} e^{\frac{-d}{\lambda}}}, \tag{A.6}$$

where $\overline{P} \equiv \frac{1}{d}\int_0^d P(z)dz$ is the average polarization.

Thus, the net depolarizing field that takes into account the effect of screening is given by

$$E_{ds} = E_d + E_s = \frac{\overline{P}R(z) - P(z)}{\varepsilon_f}, \quad (A.7)$$

where

$$R(z) = \frac{1 + e^{\frac{-d}{\lambda}} - e^{\frac{-z}{\lambda}} - e^{\frac{z-d}{\lambda}}}{1 + e^{\frac{-d}{\lambda}} - \frac{2\lambda}{d} + \frac{2\lambda}{d}e^{\frac{-d}{\lambda}}}. \quad (A.8)$$

This depolarizing field enters the free energy functional given by eq. (1).

## Acknowledgements

This work was supported by the National Science Foundation (Grant No. DMR-0906443), the Nanoelectronics Research Initiative of the Semiconductor Research Corporation, the Materials Research Science and Engineering Center (NSF Grant No. DMR-0820521), and the Nebraska Research Initiative. Computations were performed utilizing the Research Computing Facility of the University of Nebraska-Lincoln.